\documentclass{emulateapj}

\usepackage{float}
\usepackage{amsmath}
\usepackage{epsfig,floatflt}
\usepackage{subfigure}

\received{March 1, 2004}
\accepted{May 20, 2004}
\begin{document}

\title{On Foreground Removal from the \emph{Wilkinson Microwave
Anisotropy Probe} Data by an Internal Linear Combination
Method: Limitations and Implications}

\author{H.\ K.\ Eriksen\altaffilmark{1,2,3}} \affil{Institute of
Theoretical Astrophysics, University of Oslo, P.O.\ Box 1029 Blindern,
\\ N-0315 Oslo, Norway}
\altaffiltext{1}{Also at Centre of Mathematics for Applications,
University of Oslo, P.O.\ Box 1053 Blindern, N-0316 Oslo}
\altaffiltext{2}{Also at Jet Propulsion Laboratory, M/S 169/327, 4800
Oak Grove Drive,  Pasadena CA 91109} 
\altaffiltext{3}{Also at California Institute of Technology, Pasadena, CA
  91125}

\email{h.k.k.eriksen@astro.uio.no}

\author{A.\ J.\ Banday}
\affil{Max-Planck-Institut f\"ur Astrophysik, Karl-Schwarzschild-Str.\
1, Postfach 1317,\\D-85741 Garching bei M\"unchen, Germany} 
\email{banday@mpa-garching.mpg.de}

\author{K.\ M.\ G\'orski\altaffilmark{3}} \affil{Jet Propulsion Laboratory, M/S
169/327, 4800 Oak Grove Drive, Pasadena CA 91109\\Warsaw University
Observatory, Aleje Ujazdowskie 4, 00-478 Warszawa, Poland} 
\email{krzysztof.m.gorski@jpl.nasa.gov}

\and

\author{P.\ B.\ Lilje\altaffilmark{1}} \affil{Institute of Theoretical
Astrophysics, University of Oslo, P.O.\ Box 1029 Blindern, \\N-0315
Oslo, Norway}
\email{per.lilje@astro.uio.no}


\begin{abstract}
We study the Internal Linear Combination (ILC) method presented by the
\emph{Wilkinson Microwave Anisotropy Probe} (\emph{WMAP}) science team,
with the goal of determining whether it may 
be used for cosmological purposes, as a template-free alternative to
existing foreground correction methods. We conclude that the method
does have the potential to do just that, but great care must be taken
both in implementation, and in a detailed understanding of limitations
caused by residual foregrounds which can still affect cosmological
results. As a first step we demonstrate how to compute the ILC weights
both accurately and efficiently by means of Lagrange multipliers, and
apply this method to the observed data to produce a new version of the
ILC map. This map has 12\% lower variance than the ILC map of the
\emph{WMAP} team,
primarily due to less noise.  Next we describe how to generate Monte
Carlo simulations of the ILC map, and find that these agree well with
the observed map on angular scales up to $\ell \approx 200$, using a
conservative sky cut. Finally we make two comments to the on-going
debates concerning the large-scale properties of the \emph{WMAP}
data. First, we note that the Galactic south-eastern quadrant is
associated with notably different ILC weights than the other three
quadrants, possibly indicating a foreground related
anisotropy. Second, we study the properties of the quadrupole and
octopole (amplitude, alignment and planarity), and reproduce the
previously reported results that the quadrupole and octopole are
strongly aligned and that the octopole is moderately planar. Even more
interestingly, we find that the $\ell=5$ mode is spherically symmetric at
about $3\sigma$, and that the $\ell=6$ mode is planar at the $2\sigma$
level. However, we also assess the impact of residual foregrounds on
these statistics, and find that the ILC map is not clean enough to
allow for cosmological conclusions. Alternative methods must be
developed to study these issues further.
\end{abstract}


\maketitle

\section{Introduction}

The first-year release of the \emph{Wilkinson
Microwave Anisotropy Probe} (\emph{WMAP}; Bennett et al.\
2003a) data set has presented the cosmological community with an
extraordinarily rich source of high-quality information,
allowing the constraint of specific cosmological models 
and the parameters which define them to percentage accuracy. 

Nevertheless, there remains an important goal beyond such a
statistical assessment of the Cosmic Microwave Background (CMB) sky,
namely to build an accurate image of the last-scattering surface which
captures the detailed nature and morphology of our universe, and not
simply some best-fit ensemble averaged view of it. Impediments to this
program include instrumental noise and systematic artifacts, and
foreground emission from local astrophysical objects. On a fundamental
level non-cosmological foregrounds can easily compromise any
conclusion regarding primordial physics unless properly accounted for,
while on a practical level they complicate both algorithms and
analyses. Methods for either removing, suppressing or at the very
least constraining foregrounds are therefore of great importance, and,
indeed, direct attacks on the raw data are very rarely
justified. Practically any analysis must consider sky maps which have
been processed in some way, either by explicit foreground corrections,
or by introducing a sky cut.

The importance of foreground removal has been recognized in the
community for a long time (e.g.,\ Banday \& Wolfendale 1991; Readhead
\& Lawrence 1992; Brandt et al.\ 1994; Tegmark \& Efstathiou 1996;
Tegmark 1998; Tegmark et al.\ 2000), as has the preferred method for
discriminating against such contamination, namely multi-frequency
observations. While the CMB itself contributes equally to all
frequencies (as measured in thermodynamic temperature units) due to
the black body nature of its spectrum, foregrounds are typically
strongly frequency-dependent. One may therefore distinguish between
foregrounds and genuine CMB anisotropy by studying how the signal
varies with frequency.  However, detailed subtraction of foregrounds
has generally required one of two assumptions to be made: either that
all of the physical components of the foreground emission and their
spectral behavior are known, or that accurate templates of all of the
components are available and that the appropriate spectral behavior
can be determined by fitting the templates to the available
multi-frequency data. Neither method is easily adapted to accommodate
real spatial variations in the spectral behavior of the foregrounds.

The \emph{WMAP} project appears to have systematic issues under
control, whilst considerable effort has been expended on foreground
issues, and uncertainties may still remain.  Recent detections of
non-Gaussianity (Chiang et al. 2003; Coles et al. 2004; Eriksen et
al. 2004a; Naselsky et al. 2003; Park 2004; Vielva et al. 2004), the
continuing debate over the low amplitude of large angular scale
anisotropy (see e.g., Efstathiou 2004), and a possible preferred
direction or alignment contained therein (de Oliveira-Costa et al.\
2004; Eriksen et al.\ 2004b) may yet be affected by improvements of
our ability to remove non-cosmological foregrounds.  The \emph{WMAP}
satellite observes the sky at five frequencies (23, 33, 41, 61 and 94
GHz), and using at least in part this information the \emph{WMAP} team
have applied three different methods for removing, constraining or
describing the foregrounds (Bennett et al.\ 2003b).

The first method is to produce a so-called internal linear combination
map (from now on denoted ILC), which assumes nothing about the
particular frequency dependencies or morphologies of the foregrounds.
Instead, a CMB map is reconstructed by co-adding the data at the five
frequencies (now convolved to a common angular resolution of 1 degree)
with a set of weights that minimizes the final variance of the map.
The details of the non-linear method adopted to derive these weights
have not been described by the \emph{WMAP} team. In order to
accommodate spectral variability of the foregrounds, the sky has been
partitioned into 12 separate regions and the minimum variance
criterion applied to each in turn to determine the
weights. Discontinuities between regions have been minimized by
smoothing of the weights at the boundaries. The resultant CMB map is
visually satisfactory but has complex noise properties, and indeed the
\emph{WMAP} team \emph{explicitly} warns against its use for
cosmological analysis. Nevertheless, the map has been subjected to
such studies in the literature, and indeed the \emph{WMAP} team do use
the resultant map as an input to their second foreground removal
technique.

This involves the application of a Maximum Entropy Method (MEM) in
order to construct a model of the foregrounds, component by component.
The strength of this method in principle is its ability to reconstruct
the synchrotron, free-free and dust emission and their detailed
frequency dependence on a pixel-by-pixel basis.  However, the initial
stage of the analysis must still utilize templates for these dominant
foreground components, and also establishes priors on their spectral
behavior by using the \emph{WMAP} data at the five frequencies after
correction for a CMB component as determined by the WILC method
above. As we will see later, ILC-like methods in general still allow
some leakage of foreground signal into the CMB reconstruction, and
whether this results in any feedback is difficult to determine. Again
as a consequence of complex noise properties, the resultant map has
not been considered useful for cosmological purposes.  Instead, the
\emph{WMAP} team has used the results to interpret the nature of the
foreground emission. In particular, they identify a dust-correlated
component in the lower frequency (23, 33 and 41 GHz) channels with a
spectral index of $\beta \sim -2.5$ for a spectrum of the form
$\nu^{\beta}$.  This is physically interpreted as a flat
spectrum synchrotron component in regions of star formation near the
Galactic plane, rather than to emission from spinning dust, which had
become the preferred solution to this anomalous, low frequency dust
correlated emission.  The issue remains open, but recent reanalyses by
Lagache (2003) and Finkbeiner (2004) find evidence for the latter
interpretation.  The origin of this controversy lies simply with the
fact that the component separation as implemented by \emph{WMAP} is allowed
only to produce a combined synchrotron/spinning dust solution at each
frequency, with no attempt made to separately disentangle these two
components.

The final method for foreground correction is perhaps the simplest of
the three, and it is also the preferred method for generating cleaned
maps suitable for cosmological analysis.  Rather than inherently
exploiting the frequency information contained in the data, one
subtracts external templates of the various physical components (i.e.,
maps produced by non-CMB observations made preferably at frequencies
where a specific component dominates the emission) with coupling
coefficients determined by cross-correlation with the observed maps.
This avoids the noise amplification which occurs when one co-adds the
\emph{WMAP} data alone, and has the added benefit that the resultant
maps have well-known noise properties, provided that the templates
themselves do not contribute significantly to this. Difficulties
associated with the method include uncertainties in the detailed
morphologies of the templates as scaled to the wavelengths of
interest, and the propagation of errors in the coupling coefficients
into the error budgets of specific analyses.

It should be noted that Tegmark, de Oliveira-Costa, \& Hamilton (2003)
have applied a generalization of the ILC method to the \emph{WMAP}
data. The basic idea is to allow the weighting of each map to be
scale-dependent by performing the analysis in harmonic space, the
assumption being that this allows any spatial variations in the
spectral dependence of the foregrounds to be adequately tracked.  It
is not clear to what extent real variations project onto the harmonic
eigenmodes of the analysis.  As with the ILC method, complex noise
properties result, and so it is unlikely that this method is suitable
for high precision cosmological analyses. In what follows we will
denote the map as TCM -- the Tegmark et al.\ cleaned map.

In this paper a new look is taken at the ILC method presented by
\citet{bennett:2003b}, with the main goal of determining whether a map
derived in this manner can be suitable for cosmological purposes.
Specifically, we derive a new ILC map based on Lagrange multipliers
(in what follows to be referred to as the LILC -- Lagrange Internal
Linear Combination -- map), which has 12\% lower variance than the
\emph{WMAP} science team's ILC (hereafter referred to as WILC) map. We
then generate Monte Carlo simulations 
of this map by adding white noise and foreground templates to CMB
realizations, and process these through our ILC pipeline. This allows
us to quantify the efficiency of the ILC method, and realistic
foreground residual estimates may be established.

In the final section we repeat the large-scale analysis of \citet{de
Oliveira-Costa:2004} both for our new LILC map and for the
simulations, to assess the impact of residual foregrounds on these
statistics. However, we study not only the quadrupole and the
octopole, but also consider the properties of the $\ell=4, 5$ and 6
modes. In fact, we find that the properties of the latter two are at
least as intriguing as those of the quadrupole and octopole: the $\ell=5$
mode is highly spherically symmetric, and the $\ell=6$ mode is planar.

\section{Method}
\label{sec:algorithm}

The ILC method as defined by \citet{bennett:2003b} is based on a
simple premise: suppose there are $k$ 
observed CMB maps at different frequencies (but with identical beams),
and the aim is to suppress foregrounds and noise as far as
possible. Each of the $k$ maps may be written (in thermodynamic
temperature) on the form $T(\nu_k) = T_{\textrm{CMB}} +
T_{\textrm{residual}}(\nu_k)$, where $T_{\textrm{CMB}}$ and
$T_{\textrm{residual}}(\nu_k)$ are statistically
independent. Therefore, if one now forms the linear combination
\begin{equation}
T = \sum_{i=1}^{k} w_{i} T(\nu_i),
\label{eq:lin_comb}
\end{equation}
and requires that 
\begin{equation}
\sum_{i=1}^{k} w_{i} = 1,
\label{eq:lin_const}
\end{equation}
the resulting map may be written as 
\begin{equation}
T = T_{\textrm{CMB}} + \sum_{i=1}^{k} w_{i} T_{\textrm{residual}}(\nu_i).
\end{equation}
Thus, the response to the CMB signal is always unity since it is
independent of the frequency, and the $k-1$ free weights may be chosen to
minimize the impact of the residuals. Assuming the CMB component is
statistically independent of the foregrounds and the noise, one
convenient measure for this is simply the variance of $T$,
\begin{equation}
\textrm{Var}(T) = \textrm{Var}(T_{\textrm{CMB}}) +
\textrm{Var}\biggl(\sum_{i=1}^{k} w_{i}
T_{\textrm{residual}}(\nu_i)\biggr).
\label{eq:minimization}
\end{equation}
The internal linear combination method may now be defined succinctly
in terms of Equations \ref{eq:lin_comb} and \ref{eq:lin_const}, where
the weights are determined by minimizing the variance in Equation
\ref{eq:minimization}.

We compute the ILC weights by means of Lagrange multipliers. Our
Lagrange multiplier procedure is similar to the approach taken by
\citet{tegmark:2003} for computing the harmonic space weights from
which their map is constructed. A useful review of this method is also
given by \citet{tegmark:1998}.
The variance of $T$ is seen to be a quadratic form in the
weights $w_i$, and its minimization under the constraint given in
Equation \ref{eq:lin_const} is therefore most conveniently carried out
by means of Lagrange multipliers. As shown in Appendix A the linear
system of equations to be solved can be written on the following form
\begin{equation}
\begin{bmatrix}
2\mathbf{C} & \mathbf{-1} \\ 
\mathbf{1}^{\textrm{T}} & 0
\end{bmatrix} 
\begin{bmatrix}
\mathbf{w} \\ \lambda
\end{bmatrix} 
= 
\begin{bmatrix}
\mathbf{0} \\ 1
\end{bmatrix},
\label{eq:lin_eq}
\end{equation} 
where $\lambda$ is an arbitrary constant, $\mathbf{w} = (w_{1},
\ldots, w_{k})^\mathrm{T}$ are the ILC weights, and
\begin{equation}
C_{ij} \equiv \langle \Delta T_i \Delta T_j \rangle =
\frac{1}{N_{\textrm{pix}}} \sum_{p=1}^{N_{\textrm{pix}}} (T^i(p) -
\bar T^i) (T^j(p) - \bar T^j)
\end{equation}
is the map-to-map covariance matrix. The solutions to this system are
the usual inverse covariance weights,
\begin{equation}
w_i = \frac{\sum_{j=1}^{k} C^{-1}_{ij}}{\sum_{jk} C^{-1}_{jk}}.
\end{equation}

If the foreground properties vary strongly over the sky as a result of
spatially dependent spectral indexes, then the ILC method may perform
rather poorly.  To remedy this, one may subdivide the sky into
disjoint patches, and compute independent set of weights for each
patch.  \citet{bennett:2003b} divided the full sky into twelve regions,
eleven covering the non-uniform regions of Galactic plane, and the
last one covering the Kp2 region plus the well-behaved parts of the
Galactic plane. We will study this particular partitioning more
closely in \S \ref{sec:simulations}.

\begin{figure*}

\plotone{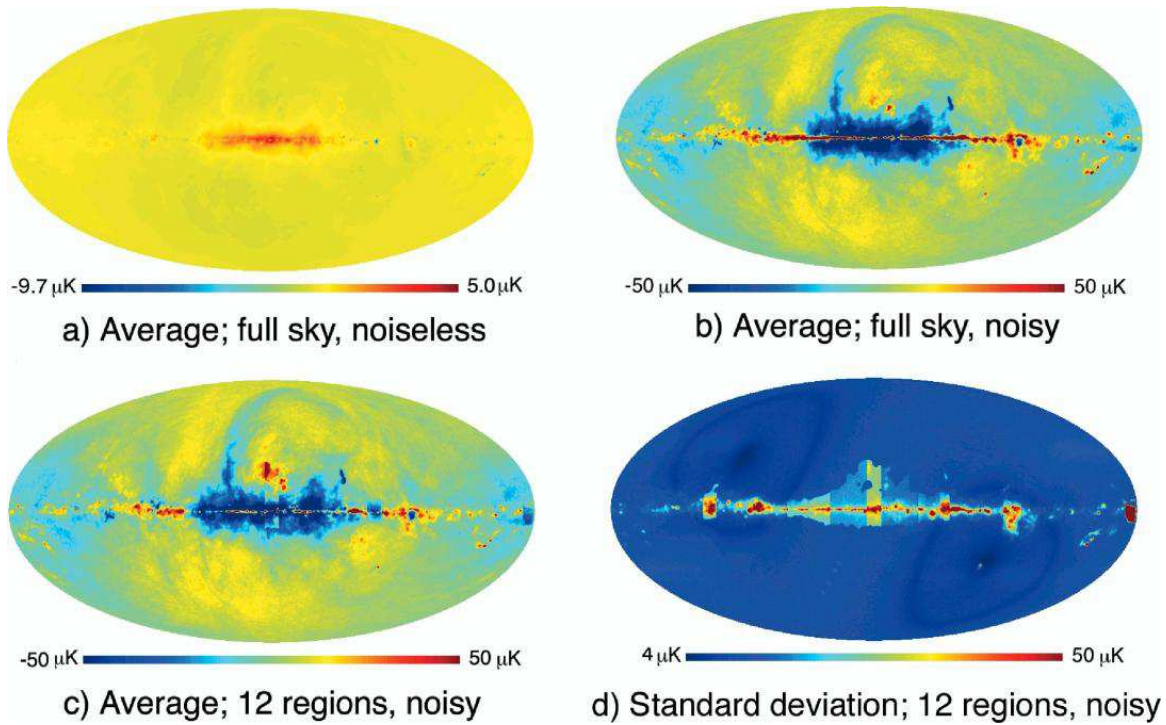}

\caption{Plots showing the pixel-by-pixel average and standard
  deviation (lower right panel) of the difference maps taken between
  the reconstructed and the input CMB maps. Each plot is computed from
  1000 simulations. \emph{Upper left:} The full sky is treated as one
  single region, and no noise is added to the simulations. \emph{Upper
  right:} Same as upper left, but \emph{WMAP} specific noise is
  included. \emph{Lower left:} The sky is partitioned into the 12
  \emph{WMAP} ILC regions, and noise is included. \emph{Lower right:} The
  standard deviation of the difference maps for which the sky is split
  into 12 regions, and noise is included.  }
\label{fig:simulated_avg}
\end{figure*}

Using such a partitioning, the minimization of the variance in
Equation \ref{eq:minimization} is carried out for each region
separately, and the final step is therefore to construct one single
full-sky map from those individual patches. In order to suppress
boundary effects \citet{bennett:2003b} generated a mask (i.e., a
full-sky map consisting of 0's and 1's) for each patch, and convolved
these masks by a Gaussian beam of $90\arcmin$ FWHM. This final ILC map
was then constructed by first generating one full-sky map from each
ILC weight set, as described above, and then they co-added these maps
pixel-by-pixel with weights given by the apodized masks. We adopt the
same method for suppressing boundary effects without modifications.

\section{Simulations, calibration and performance}
\label{sec:simulations}

Most cosmological CMB analyses are based on Monte Carlo simulations,
which in most cases is the only straightforward method of taking into
account such real-world nuisances as non-uniformly distributed noise,
non-Gaussian beam profiles and complex Galactic cuts. If the ILC
cleaned map is to be used for such purposes, one must be
able to construct a Monte Carlo ensemble that reproduces the 
detailed properties of the observed map. In this section, we first 
discuss how to produce such
an ensemble, and then we take advantage of the simulations to study
the properties of the ILC method itself.

\subsection{The simulation pipeline}

Monte Carlo simulation of the ILC map amounts simply to producing a
set of $k$ base frequency maps with similar properties to the observed
data, which are then processed through the ILC pipeline. The ILC
pipeline may then in many respects be regarded simply as one among
many statistics we apply to our maps -- the crucial part is not the
ILC pipeline in itself, but the construction of the base maps. The
only difference from main-stream simulation is that we in this case
\emph{add} foregrounds to the \emph{simulations}, rather than
\emph{subtract} them from the \emph{observations}.

The simulation process may be written in the following algorithmic
form\footnote{Although summarized specifically for the \emph{WMAP}
processing, the method can clearly be generalized to other
multi-frequency experiments.}:
\begin{enumerate}
\item Simulate one CMB component for each realization based on some
  power spectrum, and convolve this with the appropriate 
  channel-specific beams.
\item Add channel-specific foreground templates. 
\item Add a channel-specific noise realization. At this stage the
  simulation comprises 10 sky maps which mimic the observed properties
  of the 10 \emph{WMAP} channels at 5 frequency bands.
\item For each channel-specific realization, deconvolve the beam, 
  and convolve to a common resolution corresponding to a Gaussian beam
  of $1^{\circ}$ FWHM.
\item Compute an average map for each frequency.
\item Apply the ILC pipeline.
\end{enumerate}

The only subtle point in this prescription is how to handle
foregrounds. Ideally we would like to have a perfect full-sky,
noiseless foreground template at each \emph{WMAP} frequency and for each
significant foreground (e.g., free-free, synchrotron and dust), but
unfortunately, no such templates are available.  We are therefore left
with a choice between two options.

First, we may use the Finkbeiner and Haslam templates
\citep{haslam:1982,fink:2004,fink:1999} for synchrotron, free-free and
dust emission\footnote{Versions of these maps as used in the
\emph{WMAP} analysis are available at http://lambda.gsfc.nasa.gov.},
together with the channel specific weights listed in Table 3 of
\citet{bennett:2003b}.  The channel specific weights are estimated
through direct fits to the observed data, and are therefore free of
any assumptions about the spectral parameters.  Moreover, this method
includes the contribution from the anomalous dust-correlated
foreground, without the necessity to resolve the spinning dust
controversy.  In practice, the weighted sum over the three templates
approximates the correct amplitudes very well.

\begin{deluxetable*}{lcccc}
\tabletypesize{\small}
\tablecaption{Efficiency as a function of region\label{tab:efficiency}} 
\tablecomments{The residual foreground levels and signal-to-noise
  ratios as a function of region size. 
  $f_s$ -- synchrotron fraction relative to the canonical contribution
  at 22.8 GHz; $f_{ff}$ -- the free-free fraction relative to 33.0 GHz;
  $f_d$ -- the dust fraction relative to 93.5 GHz. The numbers are
  computed from sets of 1000 Monte Carlo simulations.}
\tablewidth{0pt}
\tablecolumns{5}
\tablehead{Region  & $f_s$ & $f_{ff}$ & $f_d$
  & $\frac{\sigma_{\textrm{CMB}}}{\sigma_{\textrm{noise}}}$}
\startdata
Full Kp0 region       & $\phm{-}0.02\pm0.06$ & $0.20\pm0.13$ & $0.54\pm0.06$ &
$6.2\pm0.5$ \\
Full Kp2 region       & $-0.01\pm0.04$ & $0.15\pm0.09$ & $0.54\pm0.06$ &
$6.0\pm0.4$ \\
Full Kp4 region       & $-0.01\pm0.03$ & $0.14\pm0.07$ & $0.55\pm0.07$ &
$5.9\pm0.3$ \\
\emph{WMAP} ILC Kp2   & $-0.04\pm0.01$ & $0.08\pm0.03$ & $0.55\pm0.06$ &
$5.6\pm0.2$ \\
Full sky              & $-0.03\pm0.01$ & $0.10\pm0.02$ & $0.46\pm0.03$ &
$5.2\pm0.2$   
\enddata

\end{deluxetable*}
 
On the other hand, at low Galactic latitudes the templates approximate
the real sky very poorly because of the complexity of the foreground
emission and real spectral variations close to the Galactic plane,
thus if a full-sky simulation is required, they should not be trusted.
Nevertheless, for the purposes of calibration of our method, such
inaccuracies are unlikely to affect our primary conclusions concerning
the efficiency of the ILC method.  Moreover, as we will demonstrate,
in this implementation of the ILC method the inner Galactic region
will \emph{always} be strongly polluted by foregrounds, and should not
be used for cosmological analysis.

A second option is to use the MEM maps provided by the \emph{WMAP}
team. The advantage of this method is a much better approximation to
the true sky emission at low Galactic latitudes. Unfortunately, these
maps are intrinsically noisy, and one would therefore have to
compensate for this when adding noise to the simulations. As a result,
we adopt the simple template method in this paper, which results in
simulations having acceptable power spectra in the high-latitude
region. In fact, the simulations are visually acceptable even in the
inner Galactic region, having features very much resembling those seen
in the observed ILC map.

\subsection{Sensitivity and response to noise and region definitions}

While the ILC method itself is simple to define, it is less clear 
how accurately it allows the removal of Galactic foreground emission.
To quantify this, we utilize our simulation set containing constant and
known levels of these foregrounds, reconstruct the CMB sky estimate
for each simulation via the ILC method, and compare this to the
input CMB component.

For the initial comparison, we consider the idealized case of a
full-sky noise-less analysis, including only foregrounds and CMB.  The
results from this exercise are shown in the upper left panel of Figure
\ref{fig:simulated_avg} in terms of the average residual map computed
from 1000 simulations.  In this case the ILC method does an excellent
job of removing the foregrounds, as the residuals are less than
$10\;\mu \textrm{K}$ even in the central Galactic plane, about 0.01\%
of the K-band amplitude. The remaining residual is caused by the fact
that it is possible to find a solution with slightly lower variance
than even the true solution.

The upper right panel shows the results from a similar full-sky
simulation, but for which Gaussian, channel-dependent noise has been
added to each realization. The effect is striking, indeed, as both the
Galactic plane and the North Galactic Spur are now clearly visible.
The explanation lies of course in the definition of the ILC method --
the ILC weights are defined to minimize the variance of the output
map. In the noiseless case, this is an optimization only with respect
to the foreground templates; for three templates, no variations in the
spectral indices, and four free weights to adjust, this can be
performed to high precision. However, the problem becomes more
complicated with the introduction of noise, since the minimum variance
criterion then implies a trade-off between instrument or foreground
noise. As is seen in Figure \ref{fig:simulated_avg}, a higher level of
foregrounds in a relatively small region of the sky is preferred over
increased noise over the full sky.

Obviously even the clean, high-latitude regions of the sky become
polluted by this higher level of foregrounds near the Galactic plane
when treating the full sky as one region. In order to avoid such
spreading one may therefore choose to divide the sky into separate
patches, each with rather homogeneous foreground properties, as
described above. While this procedure works very well in practice,
there are certain problems that one should be aware of.

\begin{deluxetable*}{ccccccccc}
\tablewidth{0pt}
\tabletypesize{\small}
\tablecaption{The WILC and LILC weights\label{tab:ILC_weights}} 
\tablecomments{Comparison of the official \emph{WMAP} ILC weights and the
  Lagrange multiplier weights as derived in this paper.}
\tablecolumns{12}
\tablehead{Region & Map & K-band & Ka-band & Q-band & V-band & W-band}
\startdata
1  & WILC &  \phm{-}0.10876 & -0.68367 & -0.09579 & \phm{-}1.92141 & -0.25072  \\ 
   & LILC & -0.19401 &  \phm{-}0.14004 &  \phm{-}0.07702 & \phm{-}0.61214 &  \phm{-}0.36480  \\[2mm] 
2  & WILC &  \phm{-}0.10818 & -0.67987 & -0.09017 & \phm{-}1.96859 & -0.30674  \\
   & LILC & -0.06280 & -0.14738 & -0.13982 & \phm{-}1.31073 &  \phm{-}0.03927  \\[2mm]
3  & WILC & -0.04074 & -0.28682 &  \phm{-}0.08476 & \phm{-}1.16221 &  \phm{-}0.08061  \\
   & LILC & -0.11470 &  \phm{-}0.15098 & -0.38520 & \phm{-}1.16396 &  \phm{-}0.18496  \\[2mm]
4  & WILC & -0.01847 & -0.25533 & -0.02607 & \phm{-}0.83919 &  0\phm{-}.46068  \\
   & LILC & -0.05654 & -0.01464 & -0.31223 & \phm{-}0.93407 &  \phm{-}0.44934  \\[2mm]
5  & WILC &  \phm{-}0.18610 & -0.77416 & -0.32352 & \phm{-}2.33978 & -0.42820  \\
   & LILC &  \phm{-}0.20099 & -0.86252 & -0.27825 & \phm{-}2.39309 & -0.45330  \\[2mm]
6  & WILC & -0.02158 & -0.21880 & -0.08224 & \phm{-}0.84851 &  \phm{-}0.47412  \\
   & LILC & -0.10223 &  0\phm{-}.21569 & -0.51767 & 0\phm{-}.90277 &  \phm{-}0.50144  \\[2mm]
7  & WILC &  \phm{-}0.11790 & -0.67740 & -0.09117 & \phm{-}1.94830 & -0.29763  \\
   & LILC & -0.05637 & -0.00015 & -0.45602 & \phm{-}1.46095 &  0\phm{-}.05159  \\[2mm]
8  & WILC &  \phm{-}0.12403 & -0.67639 & -0.09653 & \phm{-}1.74992 & -0.10103  \\
   & LILC &  \phm{-}0.16494 & -0.89662 &  \phm{-}0.07743 & \phm{-}2.01377 & -0.35952  \\[2mm]
9  & WILC &  \phm{-}0.10500 & -0.68438 & -0.09847 & \phm{-}1.90588 & -0.22803  \\
   & LILC & -0.04577 & -0.27660 & -0.02097 & \phm{-}1.28849 &  \phm{-}0.05484  \\[2mm]
10 & WILC &  \phm{-}0.16911 & -0.91455 & -0.01204 & \phm{-}2.64536 & -0.88788  \\
   & LILC &  \phm{-}0.19380 & -1.16103 &  \phm{-}0.37899 & \phm{-}2.26627 & -0.67803  \\[2mm]
11 & WILC &  \phm{-}0.21951 & -0.96567 & -0.18077 & \phm{-}2.38740 & -0.46046  \\
   & LILC &  \phm{-}0.22200 & -1.03357 & -0.09824 & \phm{-}2.34490 & -0.43509  \\[2mm]
12 & WILC &  \phm{-}0.11101 & -0.67501 &  \phm{-}0.05268 & \phm{-}1.59101 & -0.07970  \\
   & LILC & -0.06397 & -0.00907 & -0.46855 & \phm{-}1.92500 & -0.38342  
\enddata

\end{deluxetable*}

In the lower two panels of Figure \ref{fig:simulated_avg} we have
plotted the average (lower left panel) and standard deviation (lower
right panel) of the residual maps, when the sky is divided into the 12
regions defined by the \emph{WMAP} team. Overall the average map looks
quite similar to the full-sky case, but there are a few important
differences, namely that the inner galactic plane has a significantly
smaller amplitude, and that the blue ``halos'' around it are less
saturated. On the other hand, a few free-free regions are now visible,
which were efficiently removed when treating the entire galactic plane
as one region.

However, the two most interesting points in this respect are to be
found in the lower right plot, which shows the standard deviation of
the difference maps. First, the scanning pattern of \emph{WMAP} is clearly
visible in the high-latitude region. This indicates that noise is more
important than foregrounds in this region, and therefore the ILC
method prefers to minimize this, rather than for instance suppressing
the North Galactic Spur, which is clearly visible in the average
plot. Secondly, region number 12 (to the very right in the plot) is
associated with a very large variance and so the estimated CMB signal is
not only biased in this region, but for all practical purposes
unknown. In fact, in a number of noiseless simulations we have carried
out the ILC weight matrix is singular in this region, indicating that
there is simply too little information present here to recover the CMB
signal. When adding noise the matrix becomes non-singular, and the
procedure does yield a result, but the reconstructed field is likely
to be a very poor approximation to the underlying CMB field. The
important lesson to be drawn from this is that the size of the patches
must be large enough to provide adequate support for CMB
reconstruction. Region number 12 is too small to do this, and should
therefore either be merged into the large high-latitude region, or
extended.

\subsection{Efficiency considerations}

By assuming a fixed spectral index for each of the important
foregrounds it is possible to obtain reasonable estimates of the
residual foreground level in the ILC map. Suppose each significant
foreground may be written in the form $T_{\textrm{f}}(\nu) =
(\nu/\nu_0)^{\beta} S_0$ \citep{bennett:2003b}, where $\nu_0$ is an
arbitrary reference frequency for the particular foreground, $S_0$ is
the true foreground distribution on the sky at that frequency, $\beta$
is the spectral index and $T$ is measured in antenna temperature. Then
the residual foreground contribution to the ILC map may be written on
the form
\begin{equation}
T_{\textrm{residual}} = \biggl[\sum_{i=1}^{k} w_i \:
\biggl(\frac{\nu_i}{\nu_0}\biggr)^{\beta} a(\nu_i)\biggr] S_0 = f \cdot
S_0,  
\end{equation}
where $a(\nu)$ is the conversion factor from antenna to thermodynamic
temperature. Thus, the parameter $f$ is simply the fraction of
residual foregrounds of that particular type in the ILC map, relative
to the chosen reference frequency.

For the simulations, we know both the exact CMB component and the noise
contributions, and so we can also compute the CMB signal to noise
ratio by taking the ratio of the rms of the input CMB component to the
noise rms. The latter quantity is computed as follows
\begin{equation}
\sigma_{\textrm{noise}}^2 = \sum_{i=1}^{k} w_i^2
\sigma_{\textrm{noise},i}^{2}, 
\end{equation}
where $\sigma_{\textrm{noise},i}^{2}$ is the variance of the $i$th input
noise realization, which we know.

\begin{figure*}

\plotone{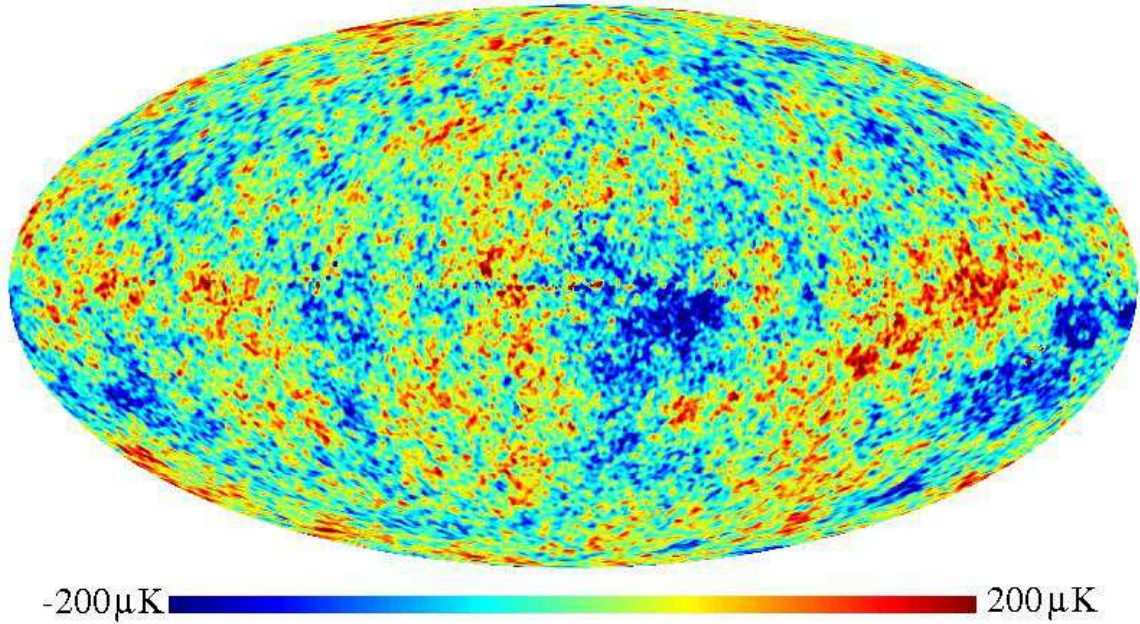}

\caption{The Lagrange Internal Linear Combination (LILC) map.}
\label{fig:myilc}
\end{figure*}

The efficiency of the ILC method may now be quantified by computing
these parameters from the Monte Carlo simulations. Such results are
summarized in Table \ref{tab:efficiency} for five different
high-latitude regions (including different amount of foregrounds). For
each quantity we list the mean and standard deviation, as computed
from a set of 1000 simulations. Three foregrounds are included here,
namely synchrotron ($\beta_{\textrm{s}} = -2.70$, $\nu_{0,\textrm{s}}
= 22.8\,\textrm{GHz}$), free-free ($\beta_{\textrm{ff}} = -2.15$,
$\nu_{0,\textrm{ff}} = 33.0\,\textrm{GHz}$) and thermal dust
($\beta_{\textrm{d}} = +2.20$, $\nu_{0,\textrm{d}} =
93.5\,\textrm{GHz}$).

The most interesting conclusions to be drawn from this table are the
following: First, the ILC method performs quite well with respect to
synchrotron emission, independently of the particular sky cut. Second,
the more area is included in the analysis, the better it does for
free-free emission, implying that the main support for information on
free-free lies close to the Galactic plane, which is a reasonable
result.

Third, the ILC method performs quite badly with respect to dust --
the residual amount of dust in the simulations is approximately half
that of the W-band, a point which must be well understood when using
the ILC map for foreground studies. We will return to this issue in
the next section.

Finally, we see that the signal-to-noise ratio increases when
excluding more of the Galactic plane. This is again a manifestation of
the competition between noise and foreground reduction. When less
foregrounds are included in the region of interest, relatively more
emphasis is put on the noise. Thus, one can easily find, somewhat
paradoxically, that by manually excluding foreground contaminated
regions from the analysis, the final amount of residual foregrounds
increases, simply because the area of choice does not carry enough
information to calibrate the ILC weights properly, and therefore the
ILC method preferentially eliminates noise rather than foregrounds. 

\section{Application to the \emph{WMAP} data}
\label{sec:application}

Table \ref{tab:ILC_weights} lists the ILC weights for each region and
for each frequency band, both as computed by \citet{bennett:2003b} and
by the Lagrange multiplier method. Figure \ref{fig:myilc} shows our
version of the ILC map.

\begin{figure*}

\plottwo{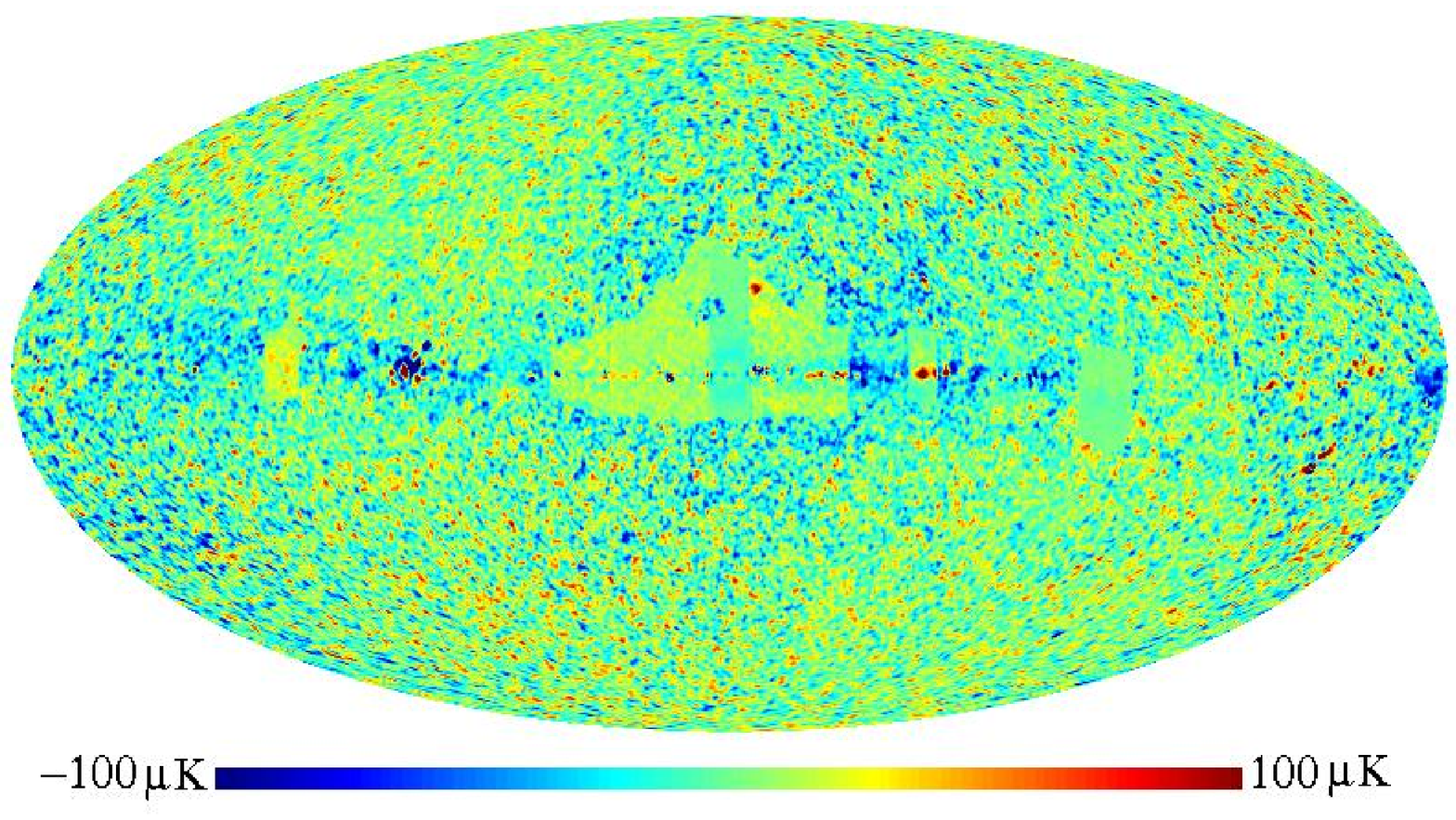}{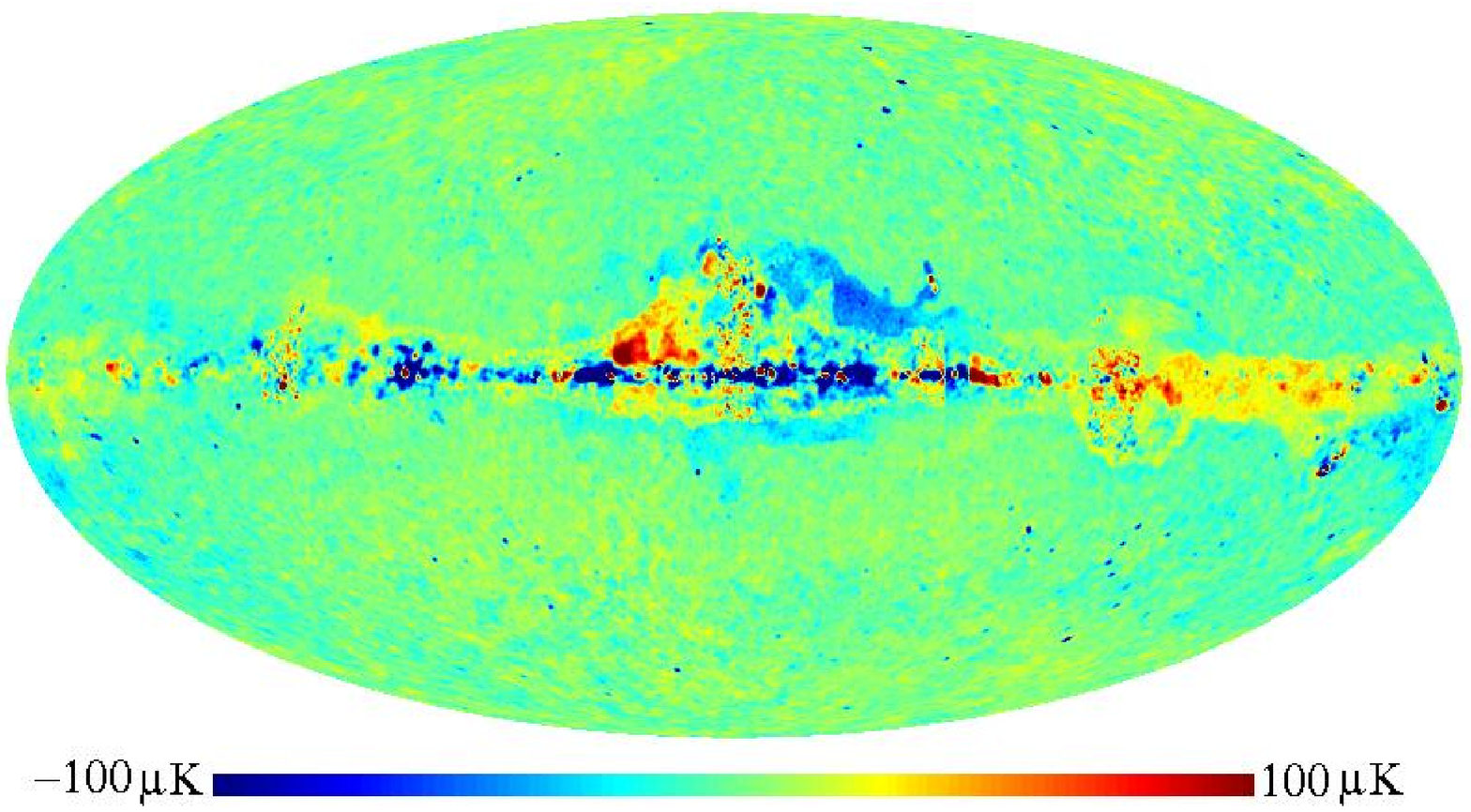}

\caption{The difference between the LILC map and a) the WILC map and b)
  the TCM. The monopole and dipole were removed from the former map, and
  the latter was smoothed to $1^{\circ}$ FWHM before differencing.}
\label{fig:diff}
\end{figure*}

A comparison of the two weight sets in Table \ref{tab:ILC_weights}
shows clearly that the differences between the two methods are
significant. The corresponding effect on the sky of these different
weights is shown in Figure \ref{fig:diff}(a), where the difference
between our map and the WILC map is plotted.  The most notable
features include the large blue area around the Galactic bulge,
presumably indicating the different abilities of the two methods to
reject some large-scale foreground structures, and secondly the
residual small-scale structure most likely indicating the different
noise properties of the two maps.

In Figure \ref{fig:diff}(b) the difference between our map and the TCM
is plotted (the TCM map was convolved to a common resolution of
$1^{\circ}$ FWHM before computing the difference).  There are no
noticeable small-scale structures uniformly distributed on the sky,
indicating similar noise properties between the two maps.  However,
there are larger scale residual foreground features present.  Some
point-source-like residuals are also present, which are associated
with known \emph{WMAP} sources.  

In order to assess the potential impact of point sources on our
method, we computed weights for the Kp2 region, both including and
excluding the 700 point sources resolved by \emph{WMAP}. The effect is
negligible, at most a one or two percent modification of the
weights. Nevertheless, this comparison does serve to remind us that
there will likely be point source residuals in any ILC-derived CMB
map.

\begin{figure}
\mbox{\epsfig{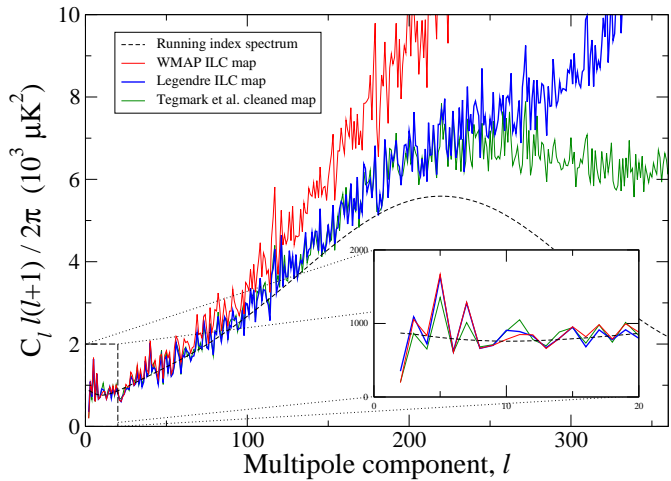}}
\caption{Comparison of the full sky power spectra of the WILC map
  (red), the LILC map (blue) and the TCM (green). Notice the excellent
  agreement between the two latter spectra up to $\ell \approx 200$,
  whereas the WILC spectrum departs from the other two already at
  $\ell \approx 50$.}
\label{fig:powspecs}
\end{figure}

Another picture of the same comparisons is given in Figure
\ref{fig:powspecs}. Here we have plotted the full-sky power spectrum
of the WILC map, the LILC map and the TCM, together with the best-fit
running index spectrum. Clearly, our map agrees very well with the TCM
up to about $\ell = 200$, but diverges at smaller scales, where the
effect of the TCM's narrower W-band beam becomes clearer. The WILC
map, however, departs from the other two already at $\ell \approx 30$,
a difference which is most naturally interpreted as resulting from
different noise properties.

\begin{figure*}
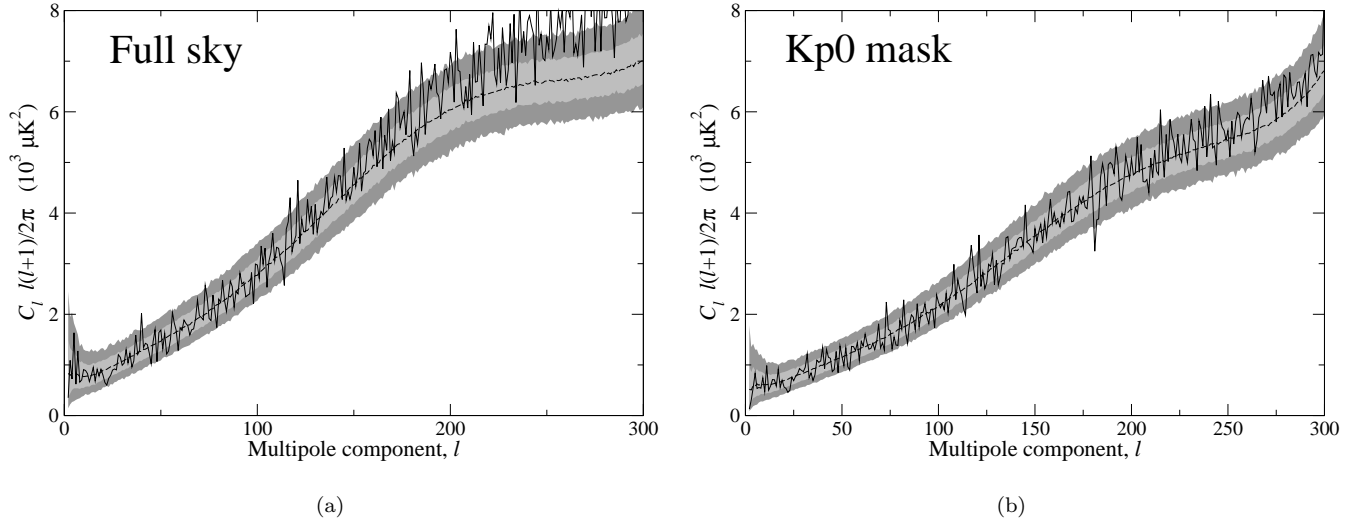


\mbox{\subfigure[]{\epsfig{figure=f5a.eps,width=0.48\textwidth,clip=}}
      \quad
      \subfigure[]{\epsfig{figure=f5b.eps,width=0.48\textwidth,clip=}}
}

\caption{Comparison of the observed LILC power spectrum to the
  simulated spectrum. The gray bands correspond to 1 and $2\sigma$
  confidence bands, estimated from 1000 Monte Carlo simulations. The
  left panel shows the spectrum computed over the full sky, and the
  right panel shows the pseudo-spectrum of the maps when the Kp0 mask
  is applied.}
\label{fig:sim_powspec}
\end{figure*}

We now compare the
observed LILC power spectrum to simulated spectra. Figure
\ref{fig:sim_powspec} shows the power spectrum of the observed ILC map
together with 1 and $2\sigma$ confidence band computed from 1000
simulations; the spectrum in the left panel is computed from the full
sky, whereas the conservative Kp0 mask has been imposed in the right
panel so that what is shown is actually a pseudo-spectrum. In the
full-sky case, we see that the observed spectrum matches the
simulations very well up to about $\ell \approx 100$, but rises more
rapidly from about $\ell > 150$. When constrained to the Kp0 region,
the observed spectrum follows the simulations all the way up to $\ell
= 200$, after which a very small bias toward high values may be
seen. Thus, the simulations seem to approximate the real sky
satisfactory on the Kp0 region, while they underestimate the level of
residual foregrounds in the inner Galactic regions.

The defining criterion of the ILC method is of course minimum
variance, and the rms of the high-latitude region of the LILC is
$68\,\mu \textrm{K}$, while the corresponding number for the WILC is
$72\,\mu \textrm{K}$. In other words, our set of weights results in
12\% lower variance, and is therefore better as far as the minimum
variance definition is concerned. However, this does not necessarily
mean that the level of residual foregrounds is smaller. In this, the
contrary is true: by computing the residual fractions of each
foreground in the high-latitude region as described in the previous
section, we find that our map actually has slightly more foreground
residuals than the WILC; the fractional residual foreground levels in
the high-latitude \emph{WMAP} ILC Kp2 region of the LILC map are [-0.069,
-0.011, 0.736], while for the WILC map they are [-0.027,-0.017,
0.424].

As noted in the previous section, the amount of residual dust is high
in the ILC maps -- the method is able to remove only half of the dust
present in the W-band, where the dust is the dominant foreground. This
result is thus in excellent agreement with the findings presented by
\citet{naselsky:2003}, which concludes that the cleaned maps contain
residual foregrounds which mainly originate from the W-band.

\subsection{Quadrant and hemisphere weights}
\label{sec:regvar}

\begin{deluxetable*}{lccccc}
\tablewidth{0pt}
\tabletypesize{\small}
\tablecaption{Hemisphere and quadrant weights\label{tab:regions}} 
\tablecomments{Weights computed from Galactic hemispheres and
  quadrants outside the Kp2 mask.}
\tablecolumns{11}
\tablehead{Region & K band & Ka band & Q band & V band & W band}
\startdata
Full Kp2 region     & -0.19401 & 0.14004 & \phm{-}0.07702 & 0.61214 & 0.36480
\\[3mm]  
Northern hemisphere & -0.20611 & 0.14837 & \phm{-}0.13262 & 0.55371 & 0.37140 
\\ 
Southern hemisphere & -0.18015 & 0.12169 & \phm{-}0.03213 & 0.66930 & 0.35703
\\[3mm]
North-west quadrant & -0.19451 & 0.13659 &  \phm{-}0.09579 & 0.56725 & 0.39489 \\     
North-east quadrant & -0.24447 & 0.21397 &  \phm{-}0.20529 & 0.53331 & 0.29190 \\
South-west quadrant & -0.19393 & 0.07268 &  \phm{-}0.26102 & 0.39229 & 0.46793 \\
South-east quadrant & -0.16324 & 0.17738 & -0.23334 & 1.01469 &
0.20451 
\enddata

\end{deluxetable*}

\begin{deluxetable}{lcccc}
\tablewidth{0pt}
\tablecaption{Maximum absolute quadrant weight differences \label{tab:maxdiff}} 
\tablecomments{The distribution of maximum absolute weight differences
  between any two Galactic quadrants, as computed from 1000
  simulations. The observed \emph{WMAP} values are shown in the
  right-most column. }
\tablecolumns{5}
\tablehead{Frequency  & Mean & Std.dev. & Max & \emph{WMAP}}
\startdata
K-band   & 0.064 & 0.028 & 0.181 & 0.081 \\
Ka-band  & 0.165 & 0.073 & 0.459 & 0.141 \\
Q-band   & 0.169 & 0.074 & 0.505 & 0.494 \\
V-band   & 0.157 & 0.071 & 0.502 & 0.622 \\
W-band   & 0.179 & 0.082 & 0.496 & 0.263 
\enddata

\end{deluxetable}

As pointed out earlier, one of the major weaknesses of the ILC method
is its inability to handle spatial variations in the spectral indices
of the foregrounds. To remedy this weakness \citet{bennett:2003b}
divided the sky into 12 disjoint regions, and computed one set of
weights for each region. Out of those 12 regions, 11 lie within the
Kp2 Galactic plane, while the rest of sky was treated as one single
region. In light of the asymmetries recently reported by
\citet{Eriksen:2004a}, we have partitioned the high-latitude sky yet
further, and subsequently computed weights for the Galactic hemispheres
and quadrants individually.  The results from these computations are
shown in Table \ref{tab:regions}.

We first consider the quadrant numbers (quadrants are defined by the
standard Galactic reference system.) While the NW, NE and SW quadrant
numbers are approximately internally consistent, the SE quadrant
stands out in the Q and V bands. Thus, these numbers both support and
ask question of the findings of \citet{Eriksen:2004a}. Certainly, the
earlier results are supported in the sense that there is an asymmetry
in the \emph{WMAP} data, possibly marginally aligned from north-west
to south-east. However, large differences in the weight coefficients
would be interpreted most naturally in terms of variations of the
noise and foreground properties, in apparent contradiction to the
frequency independence demonstrated both by \citet{Eriksen:2004a} and
\citet{Eriksen:2004b}. Further investigation is certainly warranted,
but it may yet be that foregrounds could play a role in explaining the
observed asymmetries.

Unfortunately, it is difficult to assess the significance of the
variations in Table \ref{tab:regions} properly, but we can make a few
rough estimates. We have generated 1000 simulated realizations, and
computed quadrant weights as described above for each of these. Then,
for each realization we find the maximum absolute difference between
any two quadrants, for each frequency. The results from this exercise
are summarized in Table \ref{tab:maxdiff}, in terms of the mean, the
standard deviation and the maximum value found in the
simulations. Note that these numbers are only meant to give a rough
idea of the shape of the distributions, and not for setting confidence
levels -- the distributions are non-Gaussian, and counting standard
deviations is therefore meaningless.  Nevertheless, it is obvious that
the quadrant differences observed in the true \emph{WMAP} data are
inconsistent with the adopted foreground model described by the
combination of three templates and fixed spectral indexes, and as
proposed by the \emph{WMAP} team. The weights of the south-east
quadrant are radically different from those of the other three in the
Q- and V-bands; the maximum difference found in the 1000 simulations
in the V-band is about 80\% that of the observed data. Whether this
indicates a real foreground- or noise-related problem in the
south-eastern quadrant is not clear from this analysis, but it does
question the validity of treating the entire high-latitude sky as one
single region.

The hemisphere results of Table \ref{tab:regions} are by no means as
decisive as the quadrant results, as the weights are more or less
consistent with each other. However, this may very well be a
coincidence; the internal variations between the north-west and
north-east quadrants are much smaller than between the south-west and
south-east quadrants, and yet, the two corresponding averages are
rather similar.

\section{Implications for and stability of the large-scale modes}
\label{sec:science}

In this Section we consider what implications our new LILC map have
for the current debate concerning the peculiarities seen at the very
largest scales, in particular the questions of the seemingly low
quadrupole, the planar octopole and the alignment between the two, and
we establish the uncertainties connected to each of these
measurements. For these studies we adopt the statistics of \citet{de
Oliveira-Costa:2004}, and compute 1) the probability of finding a
lower quadrupole moment than the observed one, given the best-fit
\emph{WMAP} spectrum, 2) the probability of finding such a strong
alignment between the quadrupole and octopole and 3) the probability
of finding such planar multipoles as seen in the \emph{WMAP} maps. Here we
briefly define the various statistics, and refer the interested reader
to \citet{de Oliveira-Costa:2004} for details on how each quantity
actually is computed.

\subsection{Definitions of statistics}

\begin{deluxetable}{lccc}
\tablewidth{0pt}
\tablecaption{Quadrupole amplitudes\label{tab:T_sq}} 
\tablecomments{Results from the measurements of the quadrupole
  amplitude. The third column lists the probability of finding a
  \emph{lower} quadrupole than that of the corresponding map, given
  the theoretical model value shown in the first row.}
\tablecolumns{3}
\tablehead{Measurement & $\delta T^{2}_{2} (\mu \textrm{K}^2)$ &
  p-\textrm{value}}
\startdata
Best-fit running index spectrum   & 869.7 &        \\
Hinshaw et al.\ cut sky           & 123.4 & 0.018  \\
\emph{WMAP} ILC map all sky       & 195.1 & 0.048  \\
Tegmark et al.\                   & 201.6 & 0.051  \\
Efstathiou -- WILC map            & 223.0 & 0.063  \\
Efstathiou -- TCM map             & 250.0 & 0.080  \\
Legendre ILC map                  & 350.6 & 0.153  \\
Legendre ILC map (quadrants)      & 345.1 & 0.149  
\enddata

\end{deluxetable}

The first statistic is simply the multipole amplitude $\delta T_{l}^2$, which
is defined in terms of a spherical harmonics expansion of the map,
\begin{equation}
T(\hat{\mathbf{n}}) \equiv \psi(\hat{\mathbf{n}}) = \sum_{l,m} a_{lm}
Y_{lm}(\hat{\mathbf{n}}).
\end{equation}
The multipole amplitude is then defined as 
\begin{equation}
\delta T_l^2 = \frac{l(l+1)}{2\pi}\frac{1}{2l+1} \sum_{m=-l}^{l} |a_{lm}|^2.
\end{equation}

\begin{figure}

\mbox{\epsfig{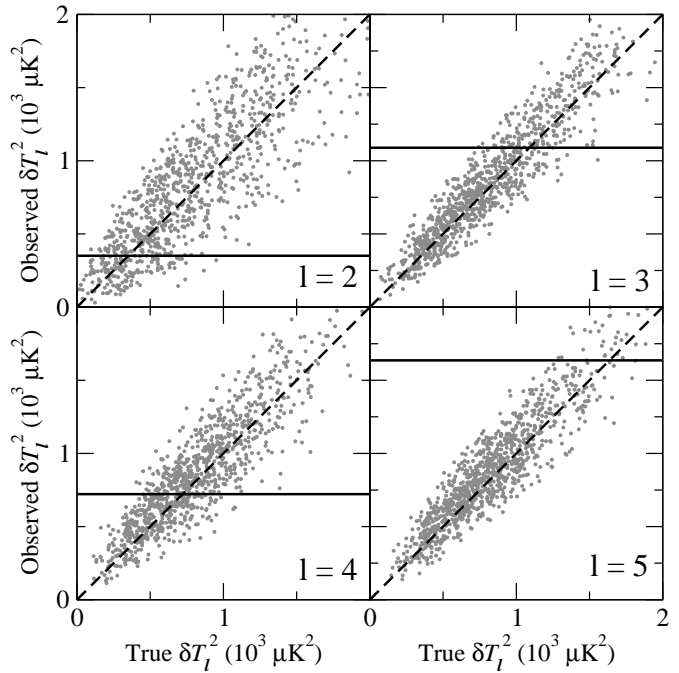}}

\caption{The observed multipole amplitude plotted against the true,
  foreground-free amplitude. The observed \emph{WMAP} LILC value is
  marked by a horizontal solid line, while the diagonal line
  is meant to guide the eye only; in the case of perfect
  reconstruction, all dots would lie along this line. Note that there
  are generally more dots above the dashed line than below it,
  indicating that the ILC reconstructed spectra are slightly biased
  toward high values.}
\label{fig:amp_error}
\end{figure}
The next statistic is based on the possibility to define a preferred
axis, $\hat{\mathbf{n}}_l$, for each multipole, namely that axis which
maximizes the angular momentum dispersion,
\begin{equation}
\bigl<\psi|(\hat{\mathbf{n}}\cdot\mathbf{L})^2|\psi\bigr> = \sum_{m=-l}^{l} m^2 |a_{lm}|^2.
\end{equation}
The alignment between two modes is then measured by taking the dot
product of the two preferred directions. 

\begin{deluxetable*}{lcccccc}
\tablewidth{0pt}
\tablecaption{Alignment of the quadrupole and the octopole\label{tab:alignment}} 
\tablecomments{Results from measurements of the position of the preferred
  directions of the quadrupole and octopole moments (denoted by
  Galactic longitude and latitude), and the alignment between
  these. The right-most column lists the probability of finding a
  weaker alignment between the quadrupole and octopole directions.}
\tablecolumns{7}
\tablehead{Measurement & $l_2$ & $b_2$ & $l_3$ & $b_3$ & Angle &
$|\mathbf{n}_2 \cdot \mathbf{n}_3|$}
\startdata
\emph{WMAP} ILC map all sky     & $278^{\circ}$ & $69^{\circ}$ & $236^{\circ}$ &
$63^{\circ}$ & $12^{\circ}$ & 0.955 \\ 
Tegmark et al.\           & $258^{\circ}$ & $59^{\circ}$ &
$238^{\circ}$ & $62^{\circ}$ & $10^{\circ}$ & 0.984 \\ 
Legendre ILC map           & $247^{\circ}$ & $62^{\circ}$ & $233^{\circ}$ &
$63^{\circ}$ & $7^{\circ}$  & 0.993 \\
Legendre ILC map (quadrants)    & $245^{\circ}$ & $61^{\circ}$ & $231^{\circ}$ &
$63^{\circ}$ & $7^{\circ}$  & 0.992 
\enddata

\end{deluxetable*}

The computation of this quantity is carried out by computing the
spherical harmonic coefficients in some coordinate system, and then
rotating these in harmonic space. Since the harmonic space rotation
matrices are simple to compute, the complete maximization procedure
becomes relatively inexpensive even for a high-resolution map with
several million pixels. The details on computing these rotation matrices are
described by \citet{de Oliveira-Costa:2004}\footnote{A complex
conjugate of the harmonic coefficients may be necessary to obtain the
same results as reported by \citet{de Oliveira-Costa:2004}, depending
on which definition of the spherical harmonics one chooses.}.

\begin{figure}

\mbox{\epsfig{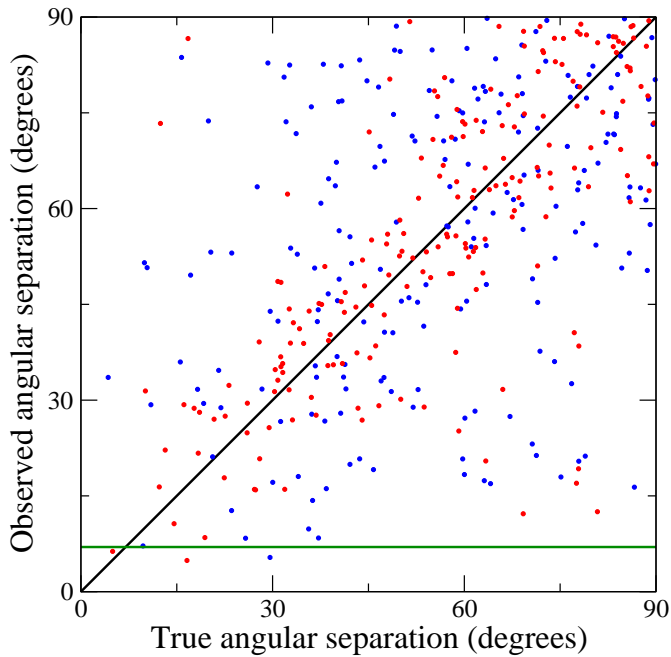}}

\caption{The observed angular separation between the preferred
  quadrupole direction, $n_2$, and the preferred octopole direction,
  $n_3$, plotted against the true, foreground-free separation. The
  color of each dot 
  indicates the quadrupole amplitude of the given
  realization. For clarity we only plot those points which either lie
  in the 0--20\% interval (blue squares) or in the 80--100\% interval
  (red dots). The horizontal line indicates the
  quadrupole value for the LILC map.}
\label{fig:directions}
\end{figure}

The third quantity we consider is the degree of planarity of a given
mode. Two different statistics for this purpose are defined by
\citet{de Oliveira-Costa:2004}, one which maximizes the angular
momentum of the mode, and one which maximizes the fractional power
that may be put into an azimuthal mode. We choose the latter, which
may be written explicitly in the following form,
\begin{equation}
t = \max_{\hat{\mathbf{n}}} \frac{|a_{l-l}|^2 + |a_{ll}|^2}{\sum_{m=-l}^{l}
  |a_{lm}|^2}. 
\end{equation}
The maximization is performed over all pixels in the map.

\subsection{Results}

In Table \ref{tab:T_sq} the amplitudes of the quadrupole moments are
tabulated for four different maps: the WILC map, the TCM, our LILC
map, and finally the LILC map for which the Kp2 region is divided into
quadrants. As we can see from the numbers in Table \ref{tab:T_sq} the
LILC quadrupole is significantly larger than those observed in the
WILC map and the TCM map. In fact, according to our map, the CMB
quadrupole is low only at the 1 to 7 level, or, in other words, it is
in perfect agreement with the model.  However, these measurements are
associated with large uncertainties. It is true that there is no
estimator induced variance in these measurements, as discussed by
\citet{efstathiou:2004}, since we have access to the full sky, but we
do know that the ILC method does not remove foregrounds perfectly in
the presence of noise, and this obviously affects the large-scale
modes.

To assess the uncertainties in these measurements we once again take
advantage of our simulations, for which we know both the CMB component
and the reconstructed map, and compare the first few low-$\ell$
amplitudes for each realization. These results are shown in Figure
\ref{fig:amp_error}.  Each black dot in these plots indicates the true
vs.\ the reconstructed amplitude for one Monte Carlo realization, and
in the limit of perfect reconstruction, they should therefore all lie
along the diagonal line. However, noise and residual
foregrounds do have a significant effect on these measurements, as
seen by the considerable scatter in each panel.

\begin{deluxetable*}{lcccccc}
\tablewidth{0pt}
\tablecaption{Planarity of a few multipoles\label{tab:planar}} 
\tablecomments{Results from measurements of the degree of planarity
  of the three multipoles, $\ell = 3, 5, 6$. The left column in each
  section shows how much of the total power in the mode is
  attributable to the $a_{ll}$ component, as measured in a coordinate
  system in which the preferred direction is defined to be the
  $z$-axis. The right column shows the probability of finding a more
  planar multipole, as compared to an ensemble of 10\,000 Gaussian
  simulations.}
\tablecolumns{7}
\tablehead{& \multicolumn{2}{c}{$\ell = 3$} & \multicolumn{2}{c}{$\ell =
    5$} & \multicolumn{2}{c}{$\ell = 6$}\\ Data set & $t$ &
  $P$ & $t$ & $P$ & $t$ & $P$}
\startdata
\emph{WMAP} ILC map           & 0.930 & 0.124 & 0.366 & 0.999 & 0.769 & 0.031 \\ 
Tegmark et al.\               & 0.942 & 0.096 & 0.372 & 0.998 & 0.783 & 0.024 \\ 
Legendre ILC map              & 0.934 & 0.114 & 0.374 & 0.998 & 0.806 & 0.015 \\
Legendre ILC map (quadrants)  & 0.948 & 0.081 & 0.375 & 0.998 & 0.794 & 0.019 
\enddata

\end{deluxetable*}

The solid horizontal lines indicate the LILC amplitudes, for which we
obviously only know the reconstructed values. In the case of the much
debated quadrupole amplitude, we see that the observed value of 351
$\mu \textrm{K}^2$ may in fact originate from a cosmological
quadrupole over the range $\sim 130$ $\mu \textrm{K}^2$ to $\sim 600$
$\mu \textrm{K}^2$, and it is therefore difficult to assign a great
deal of significance to this result. It is interesting to note that
the WILC map, which contains less residual dust than our map, also
features a lower quadrupole.  The most appropriate conclusion to draw
is that residual foregrounds can modify the quadrupole significantly,
and it is important to propagate the uncertainties in foreground
modeling into errors on such low order modes.

Given this fact, the approach taken by \cite{efstathiou:2004} might
prove more reliable if the foreground uncertainties are dominated by
residuals in the Galactic plane. In this work, the low-$\ell$ power
spectra of the WILC and the TCM maps are estimated on a cut sky (based
on the \emph{WMAP} Kp2 mask). The most likely quadrupole amplitudes
are found to be $223\mu\textrm{K}^2$ and $250\mu\textrm{K}^2$,
respectively.  An analysis by \citet{bielewicz:2004} utilizing a power
equalization filter to reconstruct the low-$\ell$ multipoles from the
high-latitude signal yields similar results.  Thus, a cut sky approach
yields slightly higher values than the corresponding full-sky analysis
does.

\begin{figure}

\mbox{\epsfig{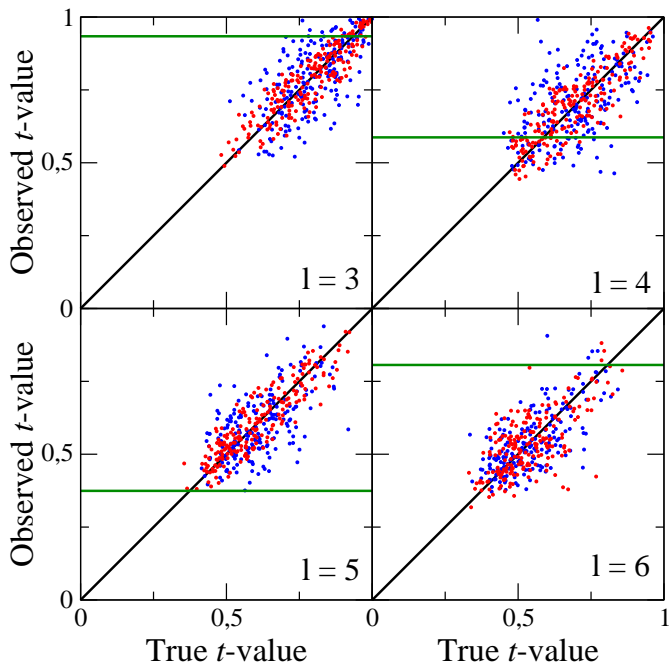}}

\caption{The observed $t$-value plotted against the true,
  foreground-free $t$-value. This parameter is defined as the fraction
  of the power attributable to the azimuthal component $a_{ll}$
  to the total power $C_{l}$, maximized over all reference
  frames. The symbols and colors have the same meanings as in Figure
  \ref{fig:directions}.}
\label{fig:t_shifts}
\end{figure}

We now turn to the question of alignment between the quadrupole and
the octopole. The results from these measurements are summarized in
Table \ref{tab:alignment}. In this case we find that the alignment is
actually stronger in the LILC map than in the WILC and TCM maps,
having a probability as low as 0.7\%. Again the associated variance is
of great importance, and a scatter plot for these measurements are
shown in Figure \ref{fig:directions}. Each dot and square in this plot
indicates the results from one simulation, for which we know both the
input and output maps. In the limit of perfect reconstruction all dots
should lie along the diagonal line. However, as seen from the very
large scatter in this plot, it is clear that the ILC method does not
reproduce the phases of the quadrupole and octopole modes accurately
enough to justify a cosmological identification.

The colors in this plot indicate the quadrupole value of the
reconstructed map for each realization, such that a realization marked
by a blue square has an amplitude smaller than 80\% of the
simulations, and a realization marked by a red dot has an amplitude
larger than 80\% of the simulations. One would expect that an
intrinsically small quadrupole is more susceptible to foreground
residuals than a strong one, and this is indeed the situation.
Further, as we have already seen, the observed LILC quadrupole value
is very low indeed, and so the conclusion of the last paragraph is
strengthened: An improved quadrupole estimate is required before we
can attach cosmological significance to its properties.  Similar
conclusions were reached in \citet{bielewicz:2004} and
\citet{hansen:2004}.

Given that the LILC map contains more residual dust than the WILC and
TCM maps and also features a stronger alignment between the quadrupole
and the octopole, one may suspect that the alignment is driven by the
dust component. However, no correlation was found between the
alignment parameter $t$ and the residual dust level
$f_{\textrm{dust}}$, or the two other foreground components. It is
therefore difficult to conclude that the alignment is a direct result
of residual foregrounds.

Despite the arguments presented in the two previous sections, it is
also worth noticing that there are very few dots below the LILC value
even for the reconstructed maps, a fact which indicates that the ILC
method does not seem to systematically introduce couplings between the
quadrupole and octopole modes. Our results therefore only demonstrate
that there is a very large variance in this measurement, but not that
there is a significant bias. The development of reliable cut-sky
estimators of this feature seem to be of high importance. 

Finally, we turn to the issue of planarity and symmetry in the low-$\ell$
modes. In Table \ref{tab:planar} we show results from measurements of
the $t$-statistic for the $\ell = 3$, 5 and 6 modes. Interestingly, as
noted by \citet{de Oliveira-Costa:2004}, the octopole is planar
roughly at the 1 to 10 level.  The $\ell=5$ and 6 modes, however, are
even more intriguing. The $\ell=5$ mode is highly spherically symmetric,
and 99.8\% of the simulations have a larger $t$-value. On the other
hand, the $\ell=6$ mode is strongly planar, with only 1.4\% of the
simulations having a larger $t$-value. For completeness, we note that
the $\ell=4$ mode appears random in all respects in our analyses.

In order to assess the foreground induced uncertainties in these
measurements, we have plotted the observed $t$-value against the true
value in figure \ref{fig:t_shifts}. It appears that the scatter
dominates the results, and it is therefore difficult to unambiguously
conclude that the detections are truly cosmological in origin.
However, we also see that the distributions are fairly symmetric about
the diagonal line, indicating that the measurements are nearly
unbiased. Residual foregrounds therefore seem to increase the variance
in the measurements, but they do not appear to introduce the sort of
effects seen in the \emph{WMAP} data. In this context, it is also important
to notice that the observed \emph{WMAP} values for $\ell=5$ and 6 are extreme
compared to that observed in the processed ILC maps. Also, the power
amplitude $\delta T^2_5$ is very large, and consequently this mode
should be quite robust against foreground perturbations.

\begin{figure*}

\plotone{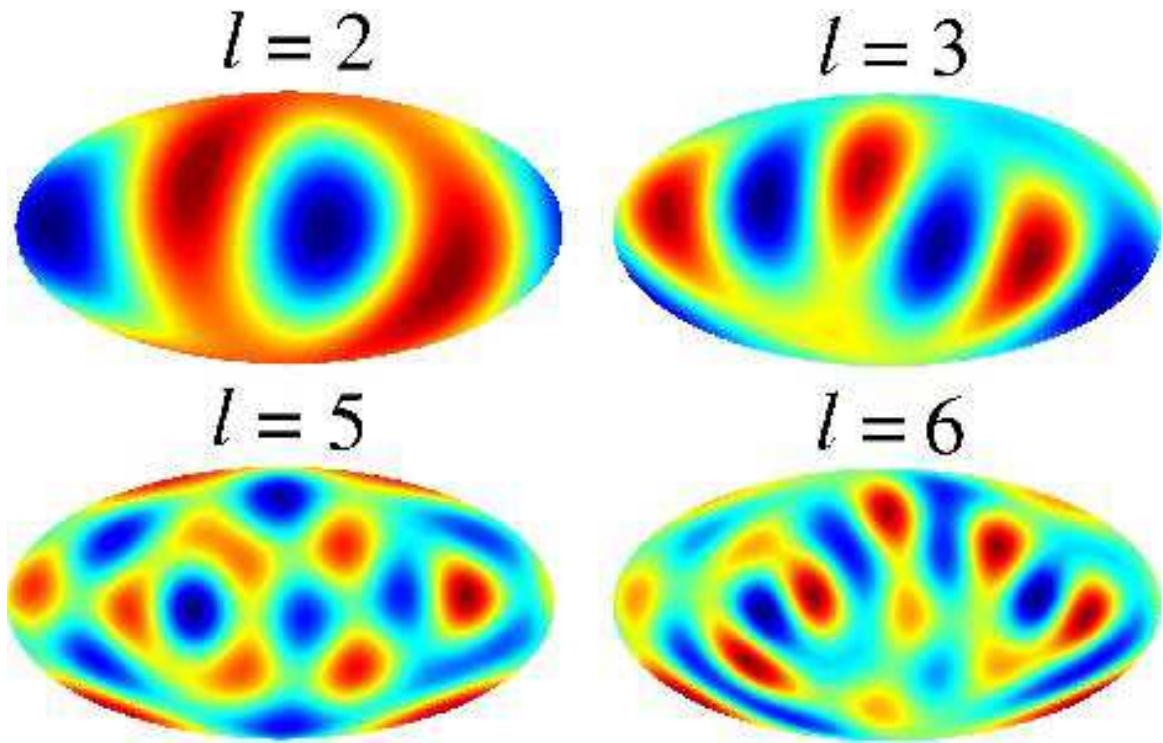}

\caption{Four peculiar low-$\ell$ modes, computed from the LILC
  map. The preferred directions of the quadrupole and the octopoles
  are strongly aligned, the $\ell = 5$ mode is spherically symmetric,
  and the $\ell = 6$ mode is planar.}
\label{fig:special_modes}
\end{figure*}

In figure \ref{fig:special_modes} we have plotted the $\ell=2, 3, 5$ and
6 modes from the LILC map. Here we clearly see the origin of the
effects discussed above: the planes determined by the peaks and
troughs of the quadrupole and octopole appear to be very strongly
aligned, while the degree of symmetry seen in $\ell=5$ is similarly
striking. Finally, the $\ell=6$ mode is obviously highly planar, as seen
by the very regular distribution of peaks and troughs. If such
features can be unambiguously shown to be of cosmological origin, they
may be indicative of new exotic physics.

\section{Conclusions}
\label{sec:conclusions}

The main goal of this paper was to study whether the ILC method is
able to yield cosmologically useful maps, and if so, whether realistic
simulations can be generated in reasonable time in order to calibrate
the uncertainties associated with the use of such a map. The results
presented earlier suggest a cautiously positive conclusion -- the ILC
method does have the capability of producing relatively clean CMB maps
without the use of external templates.  Nevertheless, great care
should be taken in the practical implementation of the method (e.g., the
proper definition of the individual regions is a crucial step), and
beyond this one needs to be highly aware of its limitations.

On a more detailed level, we derived the equations for the ILC weights
based on Lagrange multipliers, which were also discussed by Tegmark
(1998). While a non-linear search algorithm is based on iterations,
this method solves one single linear system of equations, and is
therefore much faster. This is important when generating Monte Carlo
simulations. Subsequently, we discussed how to produce realistic
simulations of the ILC map, and used these simulations to study the
properties of the method itself, with particular emphasis on the
sensitivity to noise and sky cuts.

The method was applied to the real \emph{WMAP} data, and the resultant
LILC map was determined to have properties similar to the TCM map, but
somewhat different from the \citet{bennett:2003b} WILC map. We also
computed ILC weights for four quadrants of the sky, and found that the
south-eastern Galactic quadrant has significantly different properties
than the other three, possibly shedding new light on the asymmetry
issue discussed by \citet{Eriksen:2004a}.

Finally, as a comment to the on-going debate on the nature of the
large-angular scale anisotropy, we investigated the implications of
the LILC map for estimates of the quadrupole and octopole modes, and
found that the new quadrupole moment increases from $195\;\mu
\textrm{K}$ to $351\;\mu \textrm{K}$, which is a perfectly acceptable
amplitude compared to the best-fit spectrum. However, the alignment
between the quadrupole and the octopole is stronger in our map than in
the WILC and the TCM. We also pointed out that the $\ell=5$ and 6 modes
are most peculiar in their symmetry properties, as only 0.2\% of the
simulations have a more spherically symmetric $\ell=5$ mode than the
\emph{WMAP} data, and 1.4\% a more planar $\ell=6$ mode. Further, since
we have access to the full sky, these modes are all independent under
the Gaussian, random-phase hypothesis, and the probabilities therefore
accumulate quite straightforwardly. The major caveat, however, is that
many of these measurements are derived from maps with complex
foreground and noise properties, and definitive cosmological
conclusions therefore remain elusive. Better foreground correction
methods are required, or, alternatively, methods for studying the same
properties on a cut sky should be developed. This work is already
under way, and will be published in a future paper.

Returning to the ILC method, one may question whether the minimum
variance criterion in itself is a meaningful measure of
performance. As we have seen, this criterion implies a trade-off
between suppressing noise and foregrounds, and moderate levels of
foregrounds are often accepted in order to suppress noise. For most
practical cosmological analyses this is not likely to be acceptable;
noise is more easily quantified than residual foregrounds.

Note therefore that although we do provide a copy of the LILC map at
H.K.E.'s home
page\footnote{http://www.astro.uio.no/$\sim$hke/cmbdata/WMAP\_ILC\_lagrange.fits},
we strongly advise against using it for purposes beyond visual
presentation, for which, of course, the official WILC map is perfectly
acceptable.

\begin{acknowledgements}
We thank Gary Hinshaw, Max Tegmark and Andrew Hamilton for useful
discussions, and acknowledge use of the
HEALPix\footnote{http://www.eso.org/science/healpix/} software
(G\'orski, Hivon, \& Wandelt 1999) and analysis package for deriving
the results in this paper. We also acknowledge use of the Legacy
Archive for Microwave Background Data Analysis (LAMBDA). H.\ K.\ E.\
and P.\ B.\ L.\ acknowledge financial support from the Research
Council of Norway, including a Ph.\ D.\ studentship for H.\ K.\
E. This work has also received support from The Research Council of
Norway (Programme for Supercomputing) through a grant of computing
time.

\end{acknowledgements}

\appendix

\section{Computing the ILC weights by Lagrange multipliers}
\label{app:lagrange}

In this appendix we describe how to compute the ILC weights both
efficiently and accurately by means of Lagrange
multipliers. \citet{bennett:2003b} do not specify how they carry out
the minimization of Equation \ref{eq:minimization} in practice, other
than stating that the minimum is found through a non-linear
search. Thus, it is difficult to assess the accuracy of the final
results they quote, as non-linear searches can often be plagued by
convergence issues. However, again we remind the reader that the
\emph{WMAP} team only intended their ILC map to be used for
visualization purposes, and obtaining high accuracy was therefore of
little importance. Our goal, however, was to study whether this method
may actually be used for cosmological purposes, as a credible
alternative to the template correction method. In addition, Monte
Carlo simulations was needed to fully account for the statistical
noise properties of the method, therefore computational speed was a
driving concern.

Recall that the problem is to minimize the variance of a linear
weighted sum over $k$ frequency maps, as given in Equation
\ref{eq:minimization}, under the constraint that the sum of weights
equals unity. This latter constraint guarantees a correct response to
the CMB component, while minimizing the variance suppresses residuals.

Explicitly, the variance of the final map is given by
\begin{equation}
\begin{split}
\textrm{Var}(T) &= \bigl< T^2 \bigr> - \bigl<T\bigr>^2 \\
&= \frac{1}{N_{\textrm{pix}}} \sum_{p=1}^{N_{\textrm{pix}}}
\biggl[\sum_{i=1}^{k} w_{i}T^i(p)\biggr]^2 - 
\biggl[\frac{1}{N_{\textrm{pix}}} \sum_{p=1}^{N_{\textrm{pix}}}
  \biggl(\sum_{i=1}^{k} 
  w_{i}T^i(p)\biggr)\biggr]^2\\ 
&= \sum_{i=1}^{k} \sum_{j=1}^{k} w_{i} w_{j}
\biggl[\frac{1}{N_{\textrm{pix}}} \sum_{p=1}^{N_{\textrm{pix}}}
  T^i(p) T^j(p)\biggr] - 
\biggl[\sum_{i=1}^{k} w_{i}
  \biggl(\frac{1}{N_{\textrm{pix}}}\sum_{p=1}^{N_{\textrm{pix}}}
  T^i(p)\biggr)\biggr]^2 \\ 
&= \sum_{i=1}^{k} \sum_{j=1}^{k} w_{i} w_{j}
\biggl[\frac{1}{N_{\textrm{pix}}} \sum_{p=1}^{N_{\textrm{pix}}}
  T^i(p) T^j(p) - \biggl(\frac{1}{N_{\textrm{pix}}}\sum_{p=1}^{N_{\textrm{pix}}}
  T^i(p)\biggr)^2\biggr] \\
&= \mathbf{w}^\mathrm{T} \mathbf{C} \mathbf{w}
\end{split}
\end{equation}
where $\mathbf{w} = (w_{1}, \ldots, w_{k})^\mathrm{T}$ and  
\begin{equation}
C_{ij} \equiv \langle \Delta T_i \Delta T_j \rangle = \frac{1}
{N_{\textrm{pix}}} \sum_{p=1}^{N_{\textrm{pix}}} (T^i(p) - \bar T^i)
(T^j(p) - \bar T^j)
\end{equation}
is the map-to-map covariance matrix.

Thus, the problem is simply to minimize a quadratic form, subject to
the constraint given by Equation \ref{eq:lin_const}, a task which is
most conveniently solved by Lagrange multipliers. This problem can be restated
slightly: First, we seek to minimize the following
function,
\begin{equation}
f(\mathbf{w}) = \sum_{i,j=1}^{k} w_i C_{ij} w_j
\label{eq:f_def}
\end{equation}
under the constraint 
\begin{equation}
g(\mathbf{w}) = \sum_{i=1}^{k} w_i = 1.
\label{eq:g_def}
\end{equation}
In such cases the method of Lagrange multipliers tells us to look
among those points, $\mathbf{w}_0$, which satisfies the following
relation,
\begin{equation}
\nabla f(\mathbf{w}_0) = \lambda \nabla g(\mathbf{w}_0),
\label{eq:lag_mult}
\end{equation}
where $\lambda$ is an arbitrary constant. In other words, the extrema
of $f$, subject to the constraint, $g = 1$, are just those points at
which the gradients of $f$ and $g$ are parallel.

The partial derivatives of the function $f$ are easily computed from
Equation \ref{eq:f_def}, and can be written on the following form,
\begin{equation}
\frac{\partial f}{\partial w_i} = 2 \sum_{j=1}^{k} C_{ij} w_j 
\end{equation}
The partial derivatives of $g$ are obviously just unity.

Thus, the extrema of $f$ are found by simultaneously solving the
system of $k$ derivative equations given by Equation \ref{eq:lag_mult},
and the constraint in Equation \ref{eq:lin_const}:
\begin{equation}
\begin{bmatrix}
2\mathbf{C} & \mathbf{-1} \\ 
\mathbf{1}^{\textrm{T}} & 0
\end{bmatrix} 
\begin{bmatrix}
\mathbf{w} \\ \lambda
\end{bmatrix} 
= 
\begin{bmatrix}
\mathbf{0} \\ 1
\end{bmatrix}.
\end{equation} 
It can easily be shown that the weights solving this system are
\begin{equation}
w_i = \frac{\sum_{j=1}^{k} C^{-1}_{ij}}{\sum_{jl=1}^k C^{-1}_{jl}},
\end{equation}
and so we arrive at the usual inverse covariance weighting.

\end{document}